\documentclass{desyproc}

\begin{document}

\title{Observation of diffraction and measurement of the forward energy flow with the CMS detector}

\author{{\slshape Beno\^it Roland}  on behalf of the CMS Collaboration\\[1ex]
University of Antwerp, Groenenborgerlaan 171, 2020 Antwerp, Belgium}

\contribID{xy} 
\confID{1964}
\desyproc{DESY-PROC-2010-01}
\acronym{PLHC2010}
\doi           

\maketitle

\begin{abstract}

The observation of inclusive diffraction with the CMS detector at the LHC is presented 
for centre-of-mass energies $\sqrt{s}$ = 0.9 TeV and 2.36 TeV. Diffractive events are selected 
by the presence of a Large Rapidity Gap in the forward region of the CMS detector and
uncorrected data are compared with Monte Carlo simulations based on the event generators PYTHIA and
PHOJET. The measurement of the forward energy flow, in the pseudorapidity region $3.15 < \lvert\eta\rvert < 4.9$, 
is also presented at $\sqrt{s}$ = 0.9 TeV, 2.36 TeV and 7~TeV. Uncorrected data are compared 
with Monte Carlo simulations based on PYTHIA. 

\end{abstract}

\vspace{-0.4cm}
\section{Observation of diffraction}

A diffractive reaction in $p\, p$ collisions is a reaction $p\, p \to X\, Y$ in which the systems $X$ and $Y$ 
are separated by a Large Rapidity Gap (LRG). The final states $X$ and $Y$ carry the quantum numbers of the proton 
and may be a resonance or a continuum. Diffractive reactions are described by a colourless exchange 
in the $t$ channel carrying the quantum numbers of the vacuum ~\cite{Arneodo:2005kd}. The two main types 
of diffractive processes occurring in $p\, p$ collisions are the single diffraction (SD), where one 
of the proton is scattered into a low-mass system, and the double diffraction (DD), where both protons 
dissociate. In each case the final states are characterized by an energy approximately equal to that of the incoming 
proton. Diffraction in the presence of a hard scale is described in perturbative QCD by the exchange 
of a colourless state of quarks or gluons, while soft diffraction at high energies 
is phenomenologically described in the Regge Theory~\cite{Regge:1977} by the exchange of a Pomeron.
One of the motivation to study diffraction with the early LHC data is given by the fact that a substantial fraction
of the total $p\, p$ cross section - of the order of $30 \%$ - is due to diffractive reactions, while the modelling
of soft diffraction is still mainly generator dependent. It is therefore essential to put further constraint 
on the diffractive contribution in order to improve our understanding of the collisions data 
and our knowledge of the pile up.   

\vspace{-0.2cm}
\subsection{HF calorimeter and Trigger subsystem}

A detailed description of the CMS experiment can be found elsewhere~\cite{:2008zzk} and we only describe here
the subsystems used to obtain the presented results. The two Hadronic Forward calorimeters HF+ and HF-,
located at $\pm$11.2 m from the nominal interaction point (IP), cover the pseudorapidity region 
$2.9 < |\eta| < 5.2$. These are Cerenkov calorimeters made of radiation hard quartz fibers embedded into 
steel absorbers. Half of the fibers run over the full depth of the detector, the other half start at a depth 
of 22 cm from the front face of the calorimeter. This structure enables to distinguish showers generated 
by electrons or photons from those generated by hadrons.
Two subsystems, the Beam Scintillator Counters (BSC)
and the Beam Pick-up Timing for the eXperiments (BPTX) were used to trigger the detector readout. 
The two BSCs are located at $\pm$10.86 m from the IP and cover 
the pseudorapidity region $3.23 < |\eta| < 4.65$. Each is a set of 16 scintillator tiles. The BSC elements 
have a time resolution of 3 ns and are designed to provide hit and coincidence rates. The two BPTXs, located around 
the beam pipe at $\pm$175 m from the IP, are designed to provide precise information on the bunch structure and timing
of the incoming beam, with better than 0.2 ns time resolution.  

\vspace{-0.4cm}
\subsection{Event selection}

The $p\, p$ collision data sets collected at $\sqrt{s}$ = 0.9 TeV and 2.36 TeV at the end of 2009 were used in the
analysis~\cite{CMS:diffraction}. The following conditions were imposed to select a sample with the largest acceptance 
for SD events while suppressing beam-related backgrounds. A signal is required in either of the BSCs in conjunction
with BPTX signals from both beams passing the IP. A primary vertex was required with $|z| < 15$ cm and a transverse
distance from the $z$ axis smaller than 2 cm. It was also required that at least 3 tracks be used in the vertex fitting.
Further cuts were applied to reject beam-halo event candidates and beam-scraping events. Events with large signals
consistent with noise in the hadronic calorimeter were also rejected. The energy threshold in the calorimeter was 3 GeV,
except for HF where 4 GeV was used. The number of events after the cuts are 207345 and 11848 at the two energies 
respectively~\cite{CMS:diffraction}.

\vspace{-0.4cm}
\subsection{Results}

\begin{figure}[htb]
\begin{minipage}[t]{.5\textwidth}
\centerline{\includegraphics[angle=-90,width=.9\textwidth]{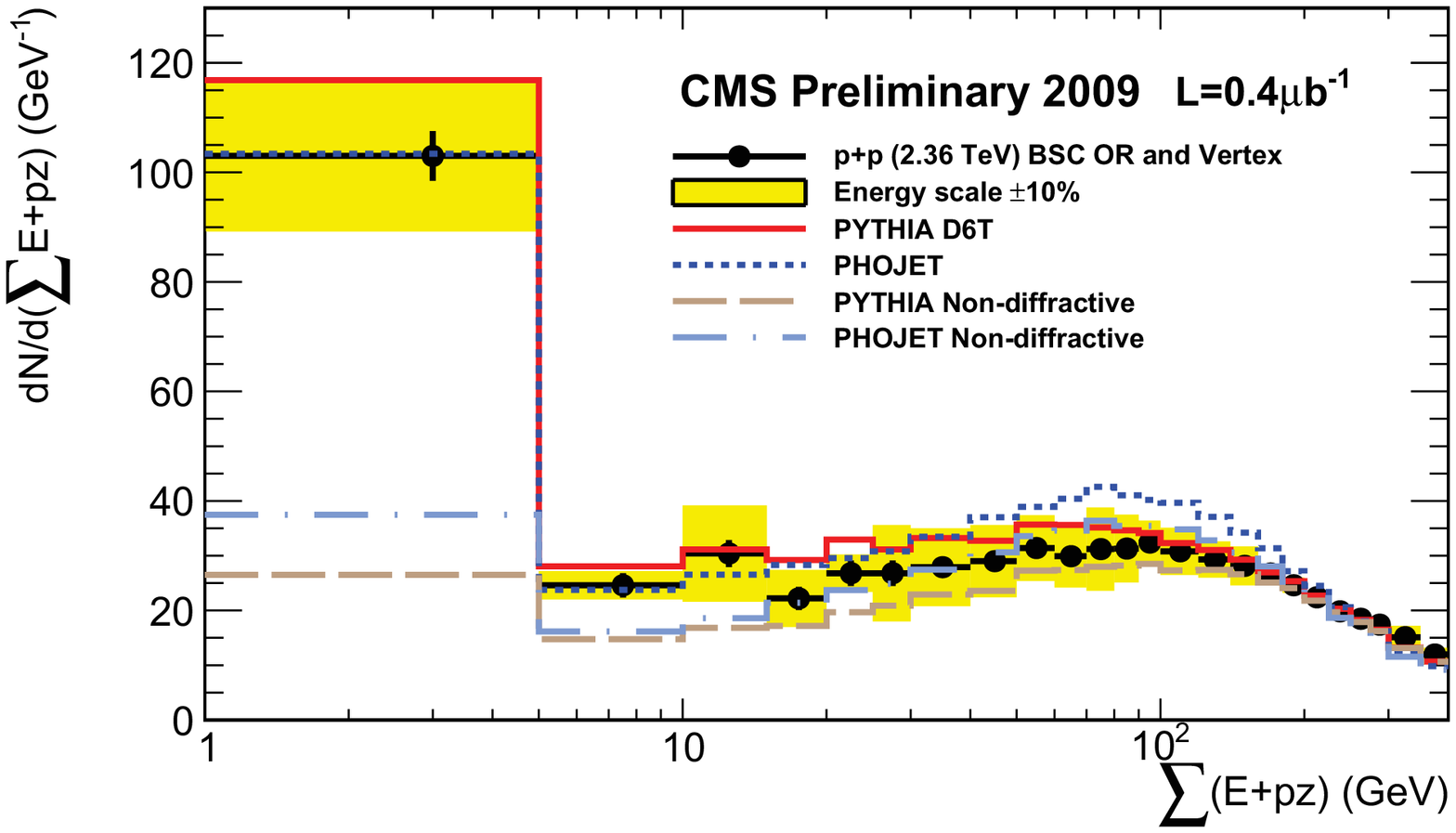}}
\end{minipage}
\begin{minipage}[t]{.5\textwidth}
\centerline{\includegraphics[angle=-90,width=.9\textwidth]{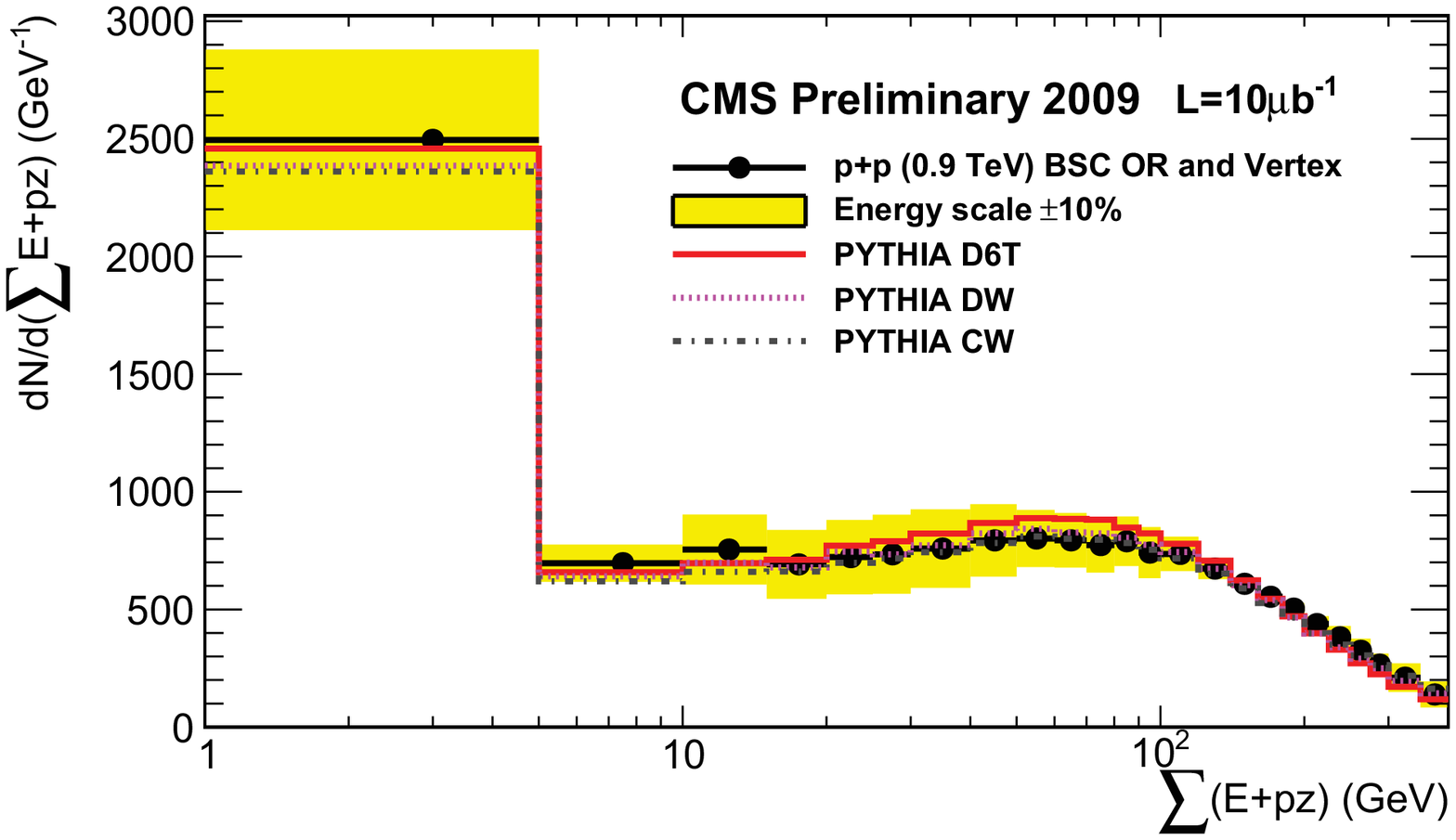}}
\end{minipage}
\begin{minipage}[t]{.5\textwidth}
\centerline{\includegraphics[angle=-90,width=.9\textwidth]{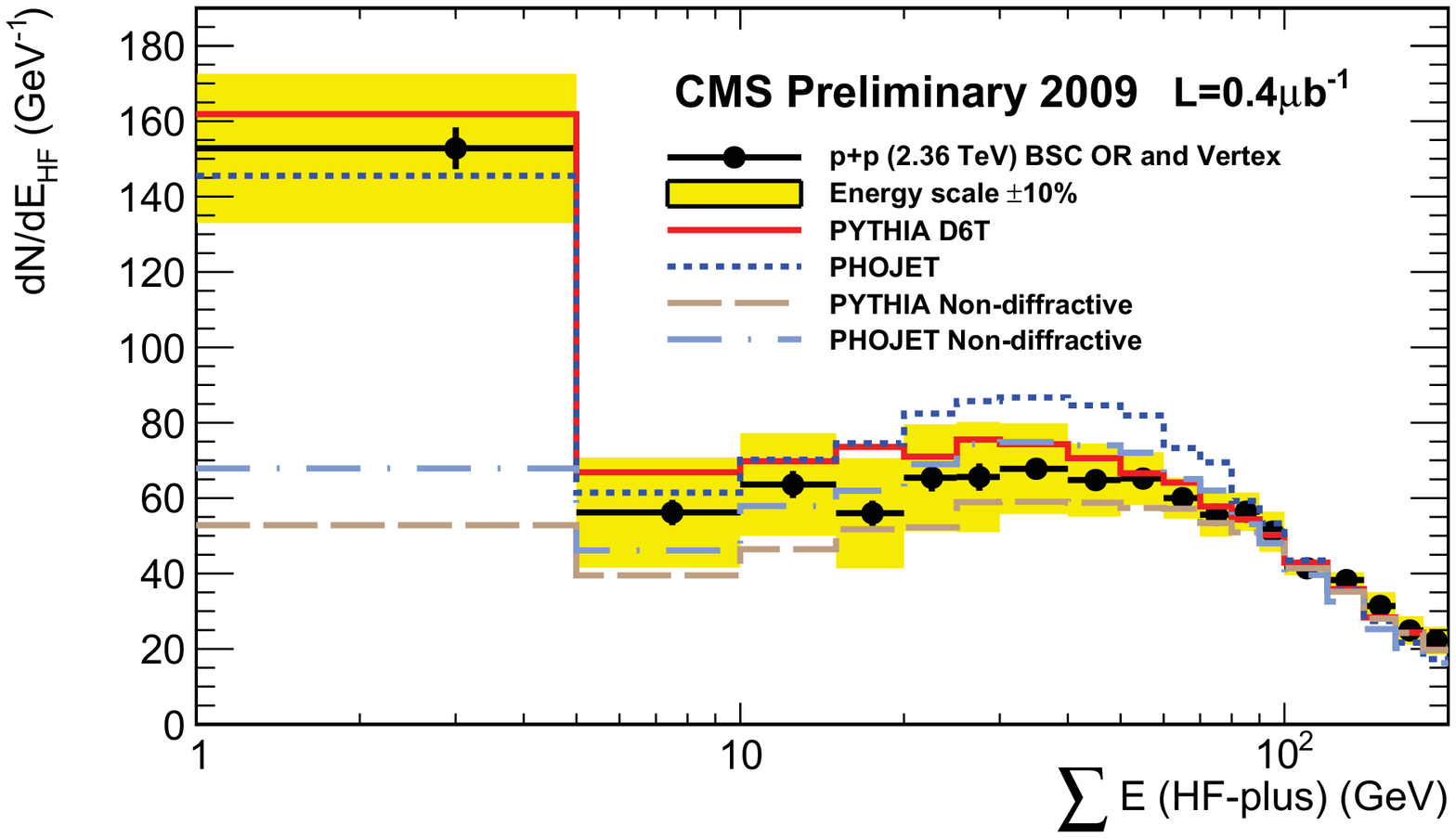}}
\end{minipage}
\hfill
\begin{minipage}[t]{.5\textwidth}
\centerline{\includegraphics[angle=-90,width=.9\textwidth]{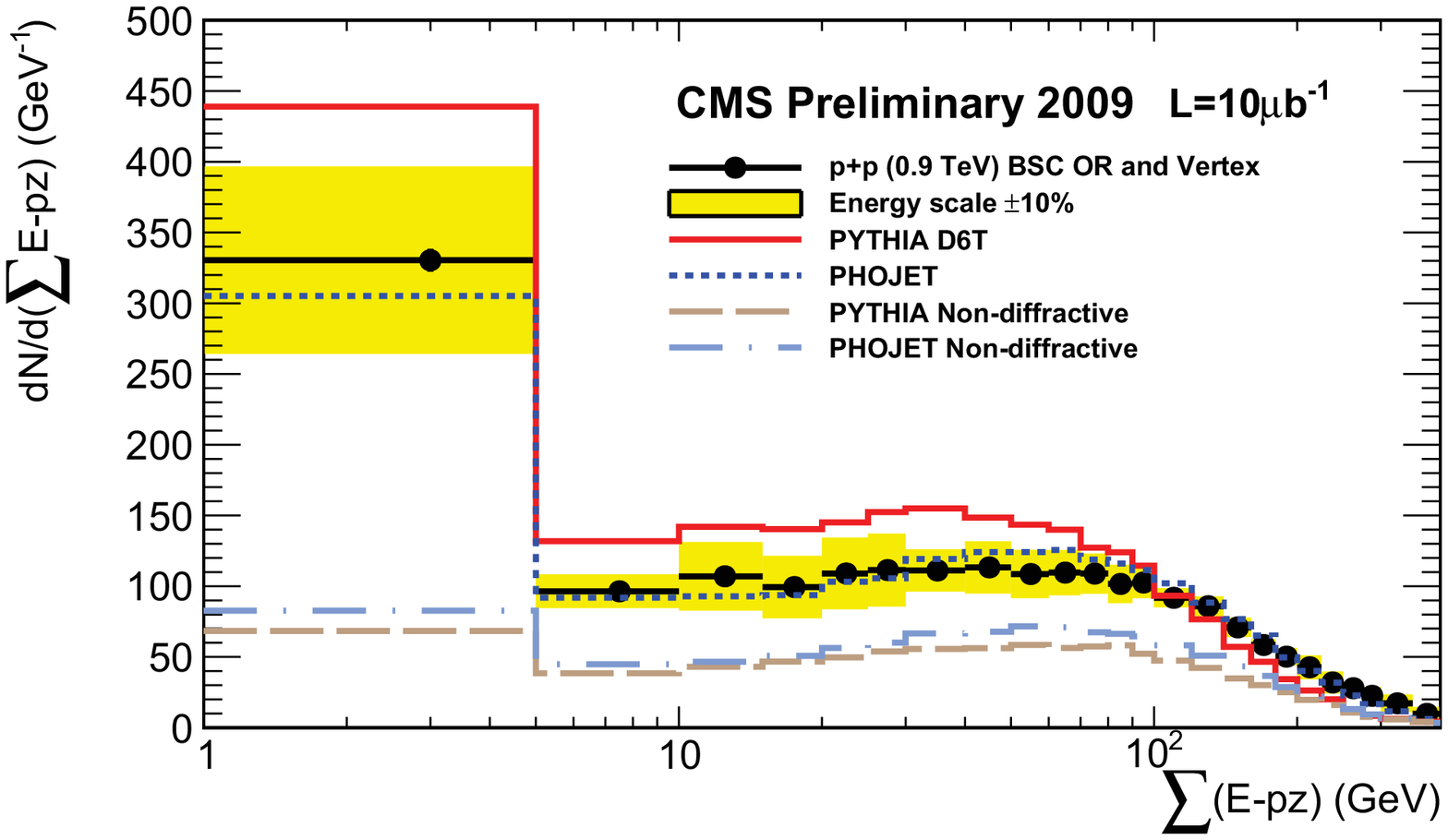}}
\end{minipage}
\hfill
\vspace{-0.5cm}
\caption{Distributions of the uncorrected variables $E + p_z$ and $E_{HF+}$ for the selected events
at $\sqrt{s}$ = 2.36 TeV compared to the PYTHIA and PHOJET predictions (left side). 
Distribution of $E + p_z$ at $\sqrt{s}$ = 0.9 TeV compared to the PYTHIA predictions 
using the tunes D6T, DW and CW (top right side). 
Distribution of $E - p_z$ at $\sqrt{s}$ = 0.9 TeV, after the requirement of $E_{HF+} < 8$ GeV, 
compared to the PYTHIA and PHOJET predictions (bottom right side).}
\label{fig:diff}
\end{figure}

The events selected at $\sqrt{s}$ = 2.36 TeV are plotted on the left side of Figure~\ref{fig:diff} 
as a function of $E + p_z$ and $E_{HF+}$. The variable $E \pm p_z = \sum_i (E_i \pm p_{z, i})$, 
where the sum runs over all calorimeter towers, approximately equals twice the Pomeron energy, with the plus (minus) 
sign applying to the case in which the proton emitting the Pomeron moves in the $+z$ ($-z$) direction. 
Diffractive events cluster at very small values of $E \pm p_z$, reflecting the peak of the cross section at small $\xi$, 
the fractional momentum loss of the proton. The variable $E_{HF+}$ represents the energy deposition in the HF+. 
Diffractive events appear as a peak in the lowest energy bin of either the HF+ or the HF-, reflecting the presence 
of a LRG extending over one of the HFs. The uncorrected data are compared to simulated events obtained from 
the PYTHIA6~\cite{Sjostrand:2006za} (tune D6T) and PHOJET 1.12-35~\cite{Bopp:1998rc,Engel:1995sb} event generators 
processed through a detailed simulation of the CMS detector based on GEANT4~\cite{Agostinelli:2002hh}.
The main systematic uncertainty is due to the imperfect knowledge of the calibration of the calorimeters and is
estimated by a 10 $\%$ variation of the energy scale. The two left side plots of Figure~\ref{fig:diff} 
show that PYTHIA gives a better description of the non-diffractive part of the data.
The events selected at $\sqrt{s}$ = 0.9 TeV are plotted on the right side of Figure~\ref{fig:diff} 
as a function of $E \pm p_z$ . The uncorrected distribution of $E + p_z$ 
is compared on the top right side to simulated events obtained from PYTHIA using the three different tunes 
D6T, DW~\cite{Bartalini:2010su} and CW~\cite{CMS:CW} for the modelling of the Multiple Parton Interactions (MPI). 
The present data can not discriminate between these different tunes. 
To enhance the diffractive component in the data, a cut was applied to the energy deposition in HF. 
The uncorrected distribution of $E - p_z$ after the requirement of $E_{HF+} < 8$ GeV is compared
on the bottom right side of Figure~\ref{fig:diff} to simulated events obtained from PYTHIA and PHOJET. 
The cut applied mainly selects SD events with a LRG extending over HF+. The system $X$ is thus boosted 
towards the $-z$ direction. The plot shows that PHOJET gives a better description of the data in this region, 
in particular of the high mass diffractive systems.
\vspace{-1.2cm}
\section{Measurement of the forward energy flow}
\begin{wrapfigure}{r}{0.6\columnwidth} 
\centerline{\includegraphics[width=0.60\textwidth,clip=]{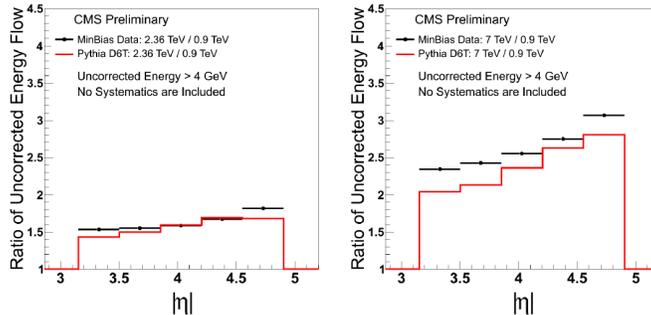}}
\vspace{-0.4cm}
\caption{Energy flow ratio as a function of $\eta$. Uncorrected data are compared to PYTHIA predictions 
using the tune D6T.}
\label{fig:eflow}
\end{wrapfigure}

Measurements in the forward region make possible to probe the small $x$ content of the proton, in a region
where the parton densities might become very large and where the probability for more than one partonic interaction
per event should increase. The measurement of the forward energy flow should therefore be sensitive to the various
modelling of the MPI~\cite{Sjostrand:1987su} and complementary to the central region measurements 
to constrain their energy dependence. 
The collision data sets collected at $\sqrt{s}$ = 0.9 TeV and 2.36 TeV at the end of 2009 
and at $\sqrt{s}$ = 7 TeV in March 2010 were used in the analysis~\cite{CMS:eflow}.
The only change wrt the selection of diffractive events is relative to the use of the BSCs.
Here a signal is required in both of the BSCs in conjunction with BPTX signals from both beams passing the IP.
The vertex requirement and the rejection of beam-related backgrounds are the same as before.
The energy flow measured in the pseudorapidity region $3.15 < |\eta| < 4.9$ is used to define 
the energy flow ratio:   
\vspace{-0.1cm}
\begin{equation}
R_{E \, flow}^{\sqrt{s_1},\sqrt{s_2}} = \frac{\frac{1}{N_{\sqrt{s_1}}} \frac{d E_{\sqrt{s_1}}}{d \eta}}
{\frac{1}{N_{\sqrt{s_2}}} \frac{d E_{\sqrt{s_2}}}{d \eta}}
\end{equation}
where $d E_{\sqrt{s}}$ is the energy deposition integrated over $\phi$ in the region $d \eta$ and $N_{\sqrt{s}}$ 
the number of selected events. The centre-of-mass energy $\sqrt{s_1}$ refers to either 2.36 TeV or 7~TeV, 
while $\sqrt{s_2}$ refers to 0.9 TeV. The two plots of Figure \ref{fig:eflow} show the ratio of the energy flows 
determined from the average of the HF+ and HF- responses~\cite{CMS:eflow}. The pseudorapidity region is divided
into 5 bins following the transverse segmentation of the HF calorimeters.
Uncorrected data without systematic uncertainties are compared to simulated events obtained from PYTHIA 
using the tune D6T~\cite{Bartalini:2010su}. While the Monte Carlo predictions agree with the data,
no conclusion on the quality of the description can be made so far due to the missing systematics. 
Figure \ref{fig:eflow} shows that the energy flow is increasing with increasing centre-of-mass energy 
and increasing $\eta$. 
\vspace{-0.7cm}
\section{Conclusion}
The observation of inclusive diffraction in $p\, p$ collisions at $\sqrt{s}$ = 0.9 TeV and 2.36 TeV
has been presented~\cite{CMS:diffraction}. Diffraction has been observed in two ways, as a peak at low
$\xi$ values and by the presence of a LRG. Uncorrected data have been compared to predictions from 
PYTHIA and PHOJET. PYTHIA describes better the non-diffractive part of the spectrum, while PHOJET gives
a better description of the diffractive system. The PYTHIA tunes D6T, DW and CW give so far a similar
description of the data. The first measurement of the forward energy flow in the HF acceptance
~\cite{CMS:eflow} ($3.15 < |\eta| < 4.9$) has been presented and compared to the PYTHIA predictions using the tune D6T.
\vspace{-0.4cm}
\section{Acknowledgments}
I would like to thank M.Arneodo, A.Vilela Pereira, S.Sen and H.Jung for usefull discussions, suggestions
and feedback.
\vspace{-0.3cm}
\section{Bibliography}

\begin{footnotesize}

\end{footnotesize}

\end{document}